\documentclass[twocolumn]{aastex631}

\begin{document}

\title{Forming Mercury from excited initial conditions}



\correspondingauthor{Jennifer Scora}
\email{jscora@sympatico.ca}

\author{Jennifer Scora}
\affil{David A. Dunlap Department of Astronomy and Astrophysics \\
University of Toronto \\
Toronto, ON, Canada, M5S 3H4}
\affil{Sidrat Research \\
124 Merton Street, Suite 507
Toronto, ON M4S 2Z2, Canada}

\author[0000-0003-3993-4030]{Diana Valencia}
\affiliation{Centre for Planetary Sciences \\
University of Toronto \\
1265 Military Trail \\
Toronto, ON, M1C 1A4, Canada}

\author{Alessandro Morbidelli}
\affil{Collège de France, CNRS, PSL Univ., Sorbonne Univ., Paris, 75014, France}
\affil{Laboratoire Lagrange, UMR7293, Universite de Nice Sophia-Antipolis, CNRS, \\
Observatoire de la Cote d'Azur. Boulevard de l'Observatoire, 06304 Nice Cedex 4, France}

\author[0000-0002-4952-9007]{Seth Jacobson}
\affil{Department of Earth and Environmental Sciences \\
Michigan State University \\
East Lansing, MI 48824, USA}



\begin{abstract}
Mercury is notoriously difficult to form in solar system simulations, due to its small mass and iron-rich composition. Smooth particle hydrodynamics simulations of collisions have found that a Mercury-like body could be formed by one or multiple giant impacts, but due to the chaotic nature of collisions it is difficult to create a scenario where such impacts will take place. Recent work has found more success forming Mercury analogues by adding additional embryos near Mercury's orbit. In this work, we aim to form Mercury by simulating the formation of the solar system in the presence of the giant planets Jupiter and Saturn. We test out the effect of an inner disk of embryos added on to the commonly-used narrow annulus of initial material. We form Mercury analogues with core-mass fractions (CMF) $> 0.4$ in $\sim 10\%$ of our simulations, and twice that number of Mercury analogues form during the formation process, but are unstable and do not last to the end of the simulations.  Mercury analogues form at similar rates for both disks with and without an inner component, and most of our Mercury analogues have lower CMF than that of Mercury, $\sim 0.7$, due to significant accretion of debris material. We suggest that a more in-depth understanding of the fraction of debris mass that is lost to collisional grinding is necessary to understand Mercury's formation, or some additional mechanism is required to stop this debris from accreting. 

\end{abstract}


\section{Introduction} \label{sec:intro}

Mercury has long been set apart from the other terrestrial solar system planets due to its orbit, diminutive size, and disproportionately large iron core. Many possible explanations have been proposed to explain its existence. The most prominent theory is that Mercury's high core-mass fraction (CMF) (0.69-0.77 \citep{2013Hauck}) and high eccentricity (0.2) originated from a collision with another planet or embryo during the giant impact phase near the end of its formation \citep{1988Benz}. The details of the collision vary from a head on, high speed impact \citep{2007Benz} to a `hit-and-run' impact where Mercury is the impactor \citep{2014Asphaug}. In the latter scenario, Mercury hits a larger body at a glancing angle, shearing off much of the mantle material from its outer edges and leaving behind a planet enriched in core material. One difficulty with this scenario is that it requires high impact velocities that are thought to be unlikely during solar system formation. Thus, it is more likely that Mercury formed from multiple subsequent collisions, allowing each collision to occur at more reasonable impact velocities \citep{2018Chau,2018Jackson,Clement2023}. Another theory for Mercury's formation is that it formed from iron-rich material that preferentially exists at, or drifts in to, the close-in semi-major axes where it formed \citep{1978Weidenschilling,2020Aguichine,2022Johansen}. \citet{2021Ryuki} suggested that the iron-rich material that forms Mercury could be generated by inner planetesimals that are stripped of their mantles as they collide with each other. This idea of Mercury forming from iron-rich material combines well with the collisional theory, as it may provide part of the iron-enrichment necessary to form Mercury, requiring fewer, or less-destructive, collisions to achieve the remaining iron-enrichment. 

Previously, it was impossible to properly test the collisional theory in simulations of solar system formation because the N-body codes used to simulate planet formation treated all collisions as perfect mergers. Multiple N-body codes have been updated over the last ten years to include more detailed collisional outcomes, including the hit-and-run necessary for one of the collisional theory's possible Mercury formation pathways. Still, some works \citep[i.e.][]{2019Lykawka,2020Fang,2022Franco} that aim to form Mercury do not make use of these codes with more detailed collisional outcomes. These updates slow simulations  down significantly, and some previous works \citep{Chambers,2016Walsh} show that, overall, the differences in the final planet configurations and accretion timescales are not statistically significant. These studies suggest that including collisional details is not worth the extra computational time required. In contrast, however, more recent works \citep[][i.e.]{2022Hagh,2022Scora} have found that detailed collisional outcomes can make an impact on dynamics and compositions of final planets, particularly in an excited environment where collisional velocities are high enough to generate significant masses of debris. More importantly, to properly test whether the collisional theory can explain Mercury's iron enrichment, collisions that are not perfect mergers are required. Thus, in this work we use an N-body code with a prescription for imperfect, debris-producing collisions to test the success of the collisional method in forming iron-rich Mercury analogues.  

Simulations forming the terrestrial planets must focus on one stage of planet formation at a time due to computational constraints. Here, we focus on the later stages of planet formation, where an initial disk of embryos grow into planets via giant impacts. The majority of previous work on this stage of formation has shown that certain conditions work best to reproduce the three larger terrestrial planets -- Venus, Earth and Mars. The initial disk of embryos produces the best Mars analogues when it begins as a truncated disk (typically between 0.7 to 1 or 1.2 AU) \citep{Hansen2009}. These truncated initial disks may have been formed due to giant planet migration (called the Grand Tack) \citep{Walsh2011,2016Walsh}, giant planet instabilities \citep{2005Tsiganis,2018Clement}, or simply because planetesimals form in these smaller rings instead of a larger disk \citep{2022Izidoro,2022Morbidelli}. Some observational evidence seems to support an early giant planet instability \citep{2023ClementI}, such as the timeline of the impacts on the Moon \citep{Morbidelli2018}, and recent simulations suggest that the dispersal of the gas disk may trigger a giant planet instability \citep{2022Liu}. However, there is some debate as to which scenario is the most in line with solar system observations.

Most of this work constraining the details of the formation of the terrestrial planets has been based on the success of the simulations in forming analogues of the three larger terrestrial planets. In some cases, Mercury's formation was ignored altogether because of the limitations of the simulations. Most often, though, it was because simulations produce Mercury analogues infrequently, and those that are produced are often too massive and too close to Venus \citep{2019Clement}. This is true of both simulations with and without prescriptions for detailed collisional outcomes \citep{2019Clement,2014Jacobson,2016Walsh,2019Lykawka}.

Recently \citet{Clement2023}, \citet{2021Clement} and \citet{2021bClement} ran simulations of Mercury's formation with collisional fragmentation that improved the occurrence rates of Mercury analogues from $< 1\%$ to $5 - 10\%$, though they remained on average too massive.  In all three cases, they added an inner disk component around Mercury's current orbit to help it form. \citet{2019Lykawka} had previously added an inner disk component without including fragmentation and similarly found that it improves the rate of Mercury analogues ($\sim 7\%$), though they did not consider Mercury's CMF.  
In this paper, we test the effect of an inner disk on the rate of forming Mercury in the presence of Jupiter and Saturn. Our treatment of debris, giant planet configurations, and inner disk profiles differ from the works of \citet{2019Lykawka,2021Clement,2021bClement,Clement2023}, and we compare our results to theirs in Section \ref{subsec:otherwork}. In Section \ref{sec:meth} we outline and justify our initial conditions and how the simulations are run. Section \ref{sec:results} and Section \ref{sec:disc} display and discuss the outcomes of these simulations, respectively. Finally, Section \ref{sec:conc} discusses the most important takeaways from this study.

\section{Methods} \label{sec:meth}

\subsection{Initial Conditions} \label{subsec:init}
\begin{table*}[]
    \centering
    \begin{tabular}{ccccc}
    \hline
      Distribution & Embryo mass & $e_{av}$ & Debris loss & Disk mass  \\
      & ($M_{\oplus}$) & & & ($M_{\oplus}$) \\
      \hline
        piecewise & 0.025 & 0.05, 0.1, 0.3 & 25\% & 3, 3.25 \\
                  & 0.025 & 0.05, 0.1, 0.3 & 50\% & 4.5, 5 \\
                  & 0.1 & 0.05, 0.1, 0.3 & 50\% & 5 \\
        \hline
        annulus & 0.025 & 0.05, 0.1, 0.3 & 25\% & 2.75 \\
        & 0.025 & 0.05, 0.1, 0.3  & 50\% & 4 \\
        \hline
    \end{tabular}
    \caption{The initial parameters for all simulations. Masses are displayed in $M_{\oplus}$.}
    \label{tab:init-params}
\end{table*}

\begin{figure}
    \centering
    \includegraphics[width=0.5\textwidth]{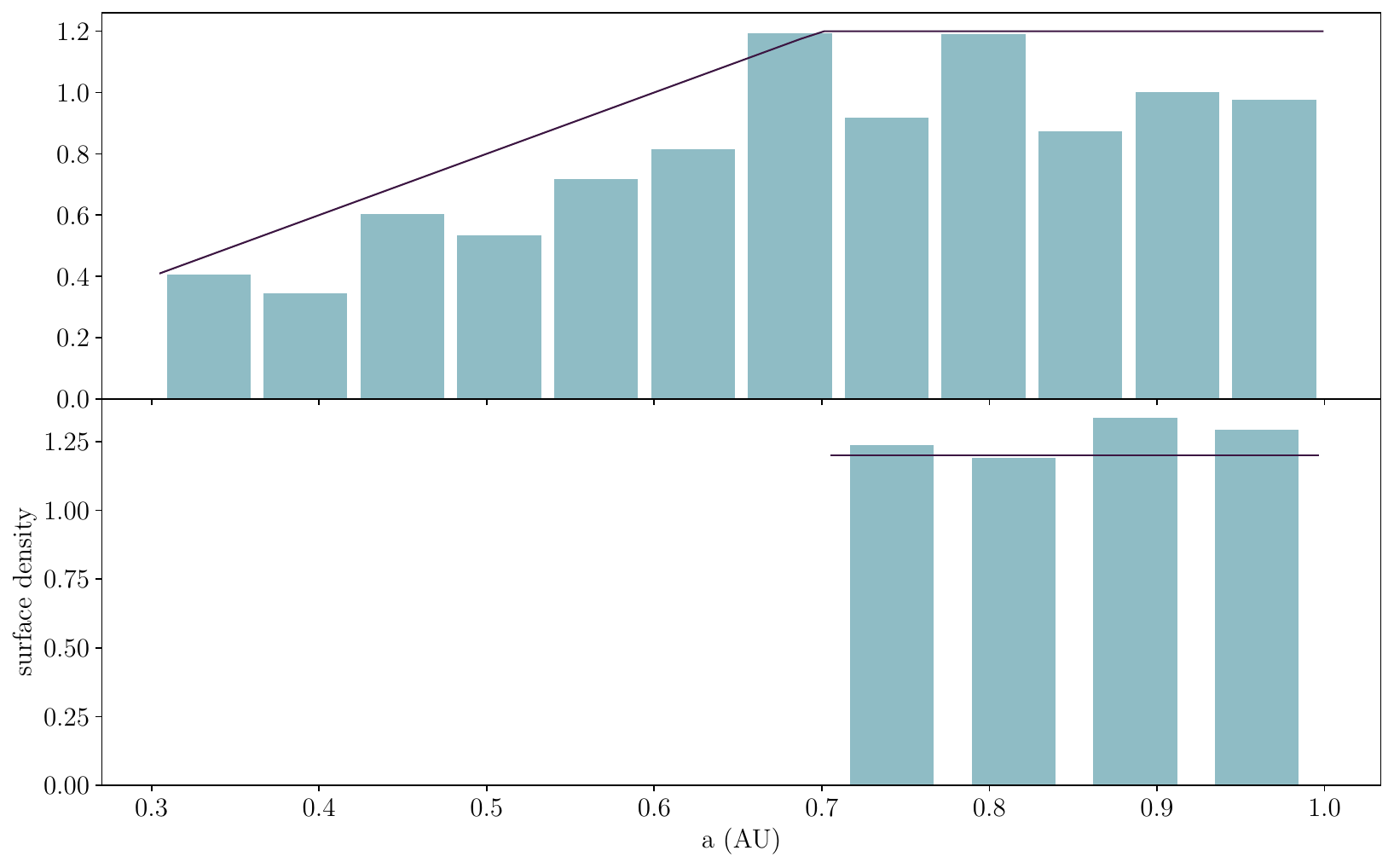}
    \caption{The surface density profiles of two runs (histogram in light blue), compared to the probability density they were drawn from (in black). \textbf{Top:} the profile for a `piecewise' distribution. \textbf{Bottom:} the profile for an `annulus' distribution.}
    \label{fig:init}
\end{figure}
To test the effectiveness of added material interior to 0.7 AU, we simulate formation starting from two different embryo distributions. The first is modelled after the successful \citet{Hansen2009} initial conditions, a truncated disk of embryos with a flat surface density between 0.7 and 1 AU (hereafter called the `annulus'). The second disk combines this initial condition with those of \citet{2021Clement} and \citet{2021bClement}, which both involve adding mass interior to 0.7 AU (hereafter called the `piecewise'). Examples of each are shown in Figure \ref{fig:init}. In particular, \citet{2021bClement} includes mass with a decreasing surface density as semi-major axis decreases, and we follow this method to include embryos with decreasing surface density inwards from 0.7 to 0.3 AU. Most simulations are run with embryo masses of 0.025 $M_{\oplus}$, and we run one set with 0.1 $M_{\oplus}$. Total disk mass is varied for the `piecewise' distribution, and is also increased with respect to the percent of debris mass that is lost to collisional grinding, as discussed in Section \ref{subsec:sims}. Table \ref{tab:init-params} lists the initial simulation parameters in more detail.

We assume an early giant planet instability. There are currently a variety of models for the time of the giant planet instability, but a general consensus is that one happened at some point before 100 Myr \citep{2018Nesvorny,2023ClementI}. We choose an early instability for three main reasons: its possible role in creating a truncated disk of embryos \citep[i.e.][]{2018Clement}, to have Jupiter's secular resonances exciting embryos forming near Mercury's orbit as in \citet{2021Clement}, and because of its potential to excite the embryos in the disk, causing more destructive, iron-enriching collisions. As the specific evolution of the giant planets during the instability is uncertain, we take a more general approach and parametrize the eccentricity excitation to explore what level of excitation is necessary to produce Mercury's high CMF. We leave to future work the assessment of whether such a level of excitation can indeed be produced by a realistic giant planet instability for the solar system.  The simulations begin after the instability has already occurred, and begin the simulations with a disk of embryos that have been excited by the instability. We randomly draw embryo eccentricities and inclinations from a Rayleigh distribution, varying the average eccentricity and inclination used (see Table \ref{tab:init-params} for the eccentricities and inclinations used).

\subsection{Simulations}\label{subsec:sims}
\begin{figure}
    \centering
    \includegraphics[width=0.5\textwidth]{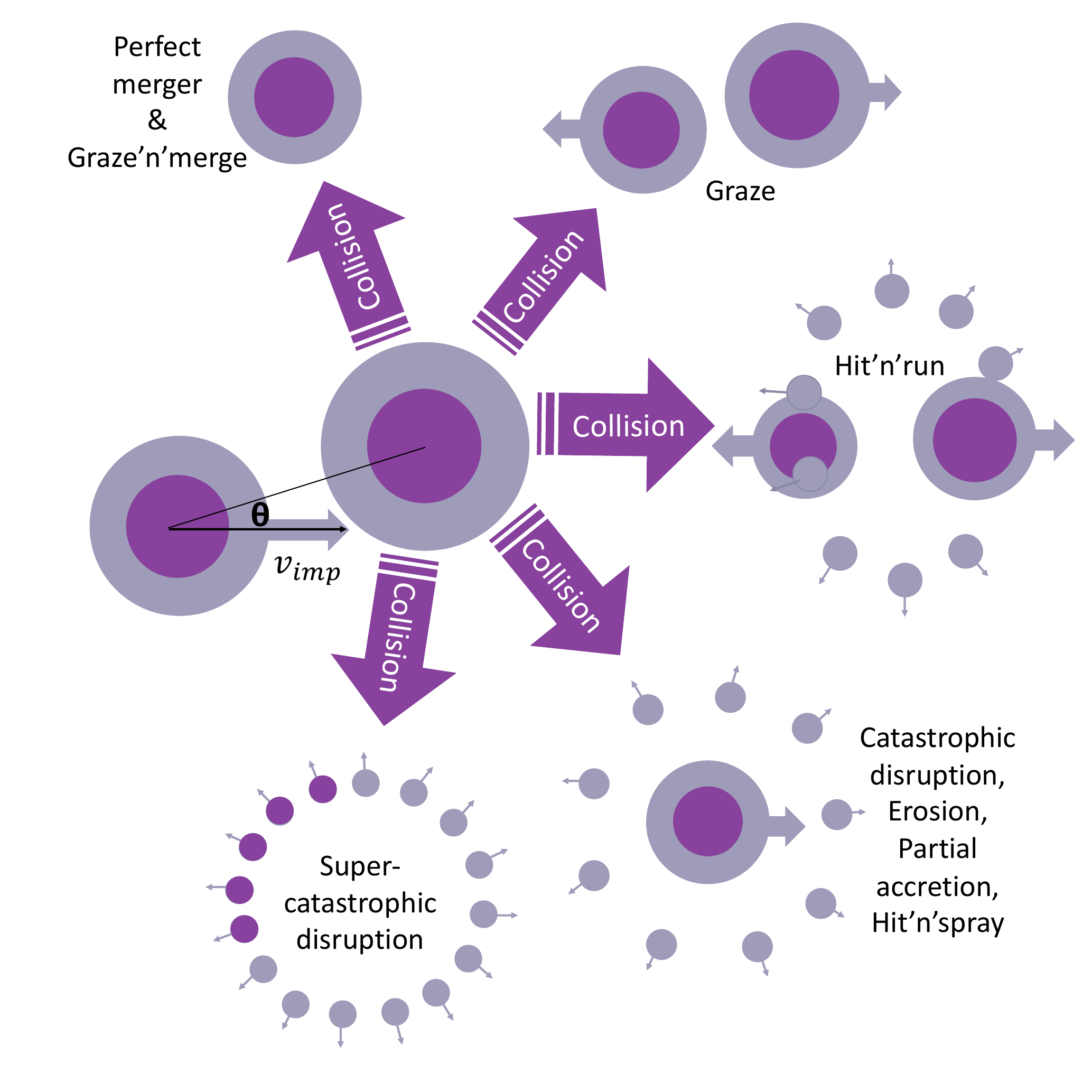}
    \caption{The nine collisional types implemented in the code are pictures here, split into their main five physical outcomes. Purple represents the core of a body, while grey represents the mantle. Large bodies are embryos, and small ones are representative of debris. }
    \label{fig:coll-graphic}
\end{figure}

The simulations begin after the dissipation of the planetary gas disk and eccentricity excitation due to the giant planet instability. The giant planets are thus assumed to be on their current orbits. We use a modified version of the gravitational N-body code \texttt{SyMBA} \citep{Duncan1998,2020Scora}. The code tracks the movement of particles within the disk based on gravitational interactions, and divides the particles into two categories: embryos ($> 2.5 \times 10^{-3} M_{\oplus}$ in these simulations) and debris. Only embryos can gravitationally interact with each other; debris particles interact with the embryos but have no gravitational effect on each other. We use a timestep of 3.65 days, and use a function to include the effects of general relativity on the orbits to improve their accuracy close to the Sun \citep{1992Saha}. 

The code uses the analytic collision prescriptions of \citet{Leinhardt2012}, \citet{Stewart}, and \citet{Genda2012} to calculate the outcomes of collisions between embryos. As depicted in Figure \ref{fig:coll-graphic}, collisions are divided into nine different types based on their impact energy, mass ratio, and impact angle. These can be subdivided into four main outcomes. Super-catastrophic disruption collisions are the most destructive, leaving behind only debris. Other disruptive collisions, such as erosive collisions, and one type of grazing collision, the hit-and-spray, leave one embryo remaining and some debris, either stripped from the target or the projectile. The other grazing collisions leave behind two embryos, and possibly some debris. Finally, the merge collisions leave behind only one embryo. When debris are generated in a collision, they are limited to a maximum of 38 particles or a minimum mass of $1.5 \times 10^{-5} M_{\oplus}$ each. After each collision where debris is created, the code removes a fraction of the total debris mass created, either $25\%$ or $50\%$. This mimics the effect of mass loss during and after the collisional process. The collisional grinding of debris into dust, and the subsequent loss of that dust as radiation pressure pushes it away from the Sun 
is the main source \citep{2012Jackson}. However, it is also possible that some of the debris mass was vapour that condensed into small, $<$ cm-sized particles that could similarly be removed via radiation pressure or other forces \citep{2007Benz}. The amount of debris mass loss chosen for this work is somewhat arbitrary, to illustrate the effect of different debris mass loss percentages on the outcomes. \citet{2020Scora} provides a more detailed breakdown of the collision treatment used in this code.

\subsection{Composition}\label{subsec:comp}

After the simulations are complete, we track the evolution of the embryos' compositions in post-processing following the method of \citet{2020Scora} and \citet{2022Scora}. Specifically, we track the embryos' core-mass fractions (CMFs), assuming an iron core and silicate mantle. We assume that all the embryos begin completely differentiated. This is based on current work that shows small bodies can differentiate in a few million years or less \citep{2021Carry,2014Neumann,2019Lichtenberg,1992Tonks}, combined with work that shows collisions enhance differentiation speed \citep{Dahl2010,Rubie,Landeau2021}, thus suggesting that our larger embryos have differentiated as they formed. For simplicity, we also assume that they have enough time to fully differentiate between collisions during our simulations, since the timescale for the cores to merge is significantly shorter than the timescale between collisions \citep{Dahl2010}.

The embryos are all assigned the same initial CMF of 0.33, based on the solar Fe/Mg abundance ratio of $2.001$ \citep{2014Palme}. Changes in the embryos' CMFs are tracked after each collision. The change in CMF depends on the type of collision and the initial CMFs of the colliding embryos. Figure \ref{fig:coll-graphic} shows the basic changes that occur for each collision type. Collisions that result in perfect mergers simply result in the adding together of all the colliding material, essentially resulting in the mass-weighted average of the CMFs of the colliding bodies. On the other end of the spectrum are the high-velocity, destructive collisions such as catastrophic disruption and erosive collisions. These are the collisions where the projectile strips material off of the outside of the target embryo, resulting in a smaller embryo than at the start. The material removed is usually mantle material, so these are the collisions with the greatest potential to increase a planet's CMF. Hit-and-run collisions can also do this, but by stripping material off of the projectile instead of the target. In general, core and mantle masses are conserved when calculating the new embryo composition. The new embryo (or embryos) are assigned their changed CMFs, and the debris is made up of the remaining core and mantle material. Core and mantle ratios are not conserved in each collisions, as a fraction of this debris material is then discarded due to debris loss. That mass is typically mantle material, since we consider that collisions strip the outer mantle material first before removing any core material. However, occasionally core material will end up in the debris, and thus sometimes core material is removed as well. More details on the specific algorithm can be found in Appendix B of \citet{2020Scora}.

\section{Results}\label{sec:results}

\subsection{Evolution of simulations}\label{subsec:evolution}

\begin{figure*}
    \centering
    \includegraphics[width=0.8\textwidth]{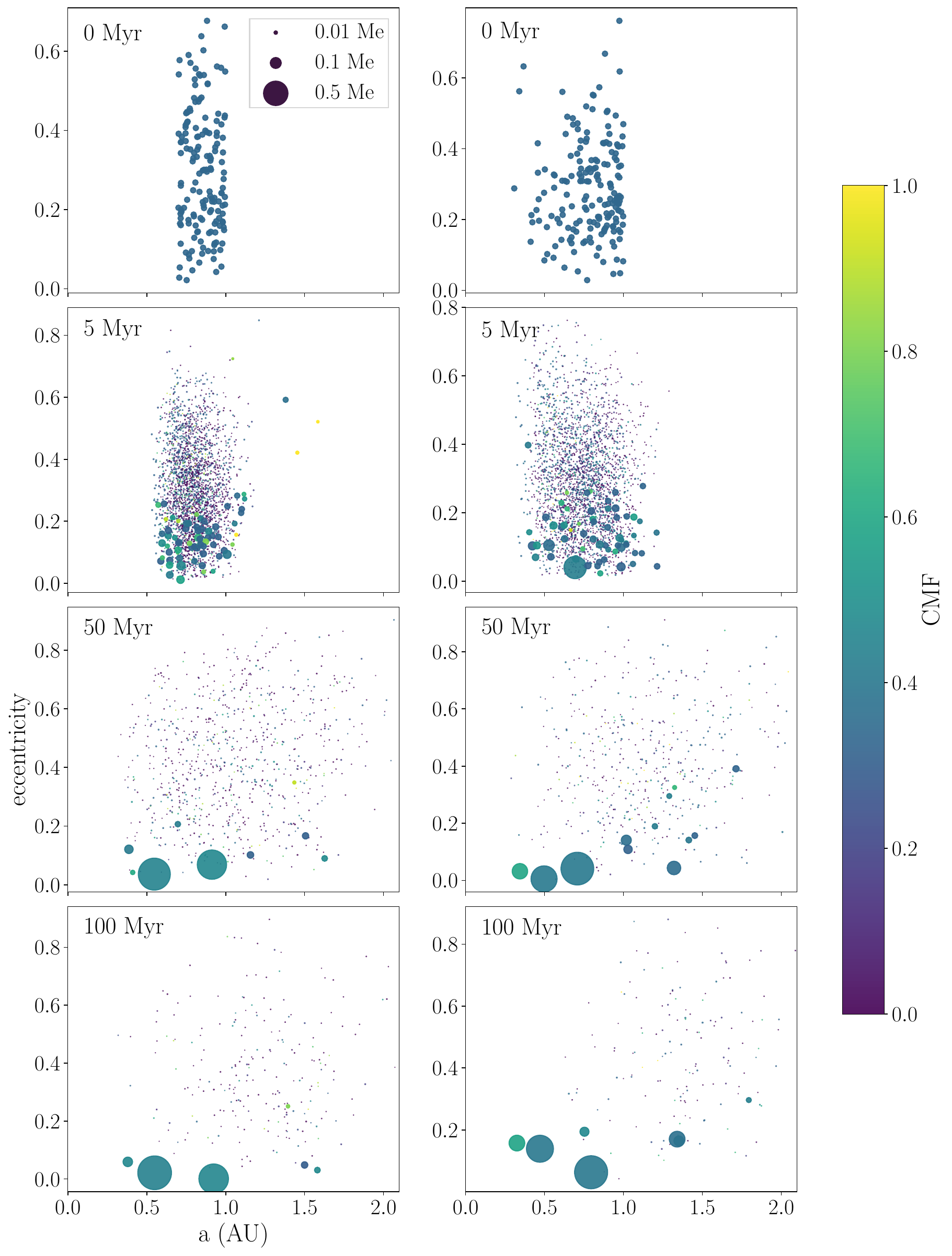}
    \caption{The evolution of sample `annulus' (left) and `piecewise' (right) simulations that form Mercury analogues. The size of the bodies represents their mass, and their colour represents their CMF.}
    \label{fig:evol-a}
\end{figure*}

We ran the simulations for 100 million years (Myrs). In that time, they typically lost $1-2 M_{\oplus}$ of mass via collisional grinding, although this left the final systems with a mass in planets $> 2 M_{\oplus}$, greater than that of our inner solar system. This is because we started with 1-2.5 times the mass needed to form the terrestrial planets, as \citet{2022Scora} found that collisions could remove more than half of the initial disk mass. Based on these new results it appears we over-corrected for the amount of debris mass loss. Figure \ref{fig:evol-a} shows the evolution of an `annulus' and `piecewise' simulation with a Mercury analogue. In both cases, the disk of embryos and debris expands outward to larger semi-major axes over time. However, in the case of the `annulus' simulation, the embryos also extend inward, while in the `piecewise' simulation the inner edge of  embryos moves outward slightly, such that the inner edge of the disk of embryos ends up quite similar for both disk types despite the initial differences. Notice that the simulations presented in the figure both start with an average eccentricity of 0.3, which lead to multiple high-CMF bodies in the intermediate stages, yet a planet with CMF $> 0.7$ survives to the final system in only one of these two simulations. 

\begin{figure}[h]
    \centering
    \includegraphics[width=0.5\textwidth]{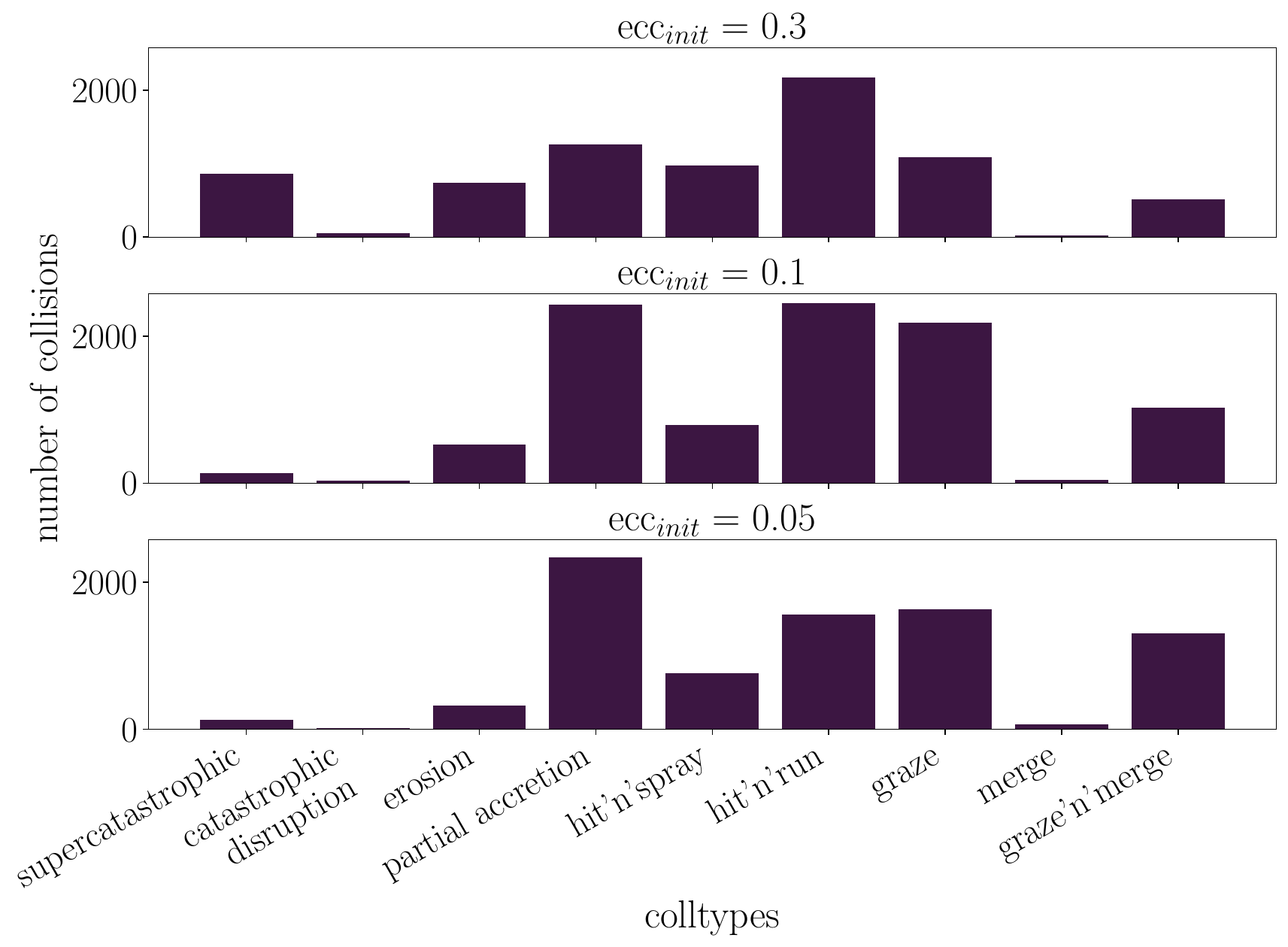}
    \caption{The collision frequencies for each collision type, for simulations with an average initial eccentricity of 0.3 (top), 0.1 (middle), and 0.05 (bottom). Hit-and-run and partial accretion collisions are most common. Erosion collisions become more common as the average initial eccentricity increases.}
    \label{fig:coll-freq}
\end{figure}

In all simulations, the mass-averaged eccentricity tends to converge on $\sim 0.15 - 0.2$ in the first few Myrs, and tends to stagnate at this level for much of the systems' evolution. Thus, the effect of the initial eccentricity is mainly confined to the first few Myrs. For those simulations that start with an average eccentricity of 0.05, this means that in the first few Myr the system has fewer collisions and those happen at much lower velocities, eventually resulting in a few embryos with larger masses. For simulations with an average initial eccentricity of 0.1, the average eccentricity only needs to increase slightly, and most of the initial collisions cause one or two embryos to get significantly more massive, dominating the rest of the evolution. For those simulations with the higher initial eccentricity of 0.3, those first collisions are very destructive, and cause a few embryos to become smaller before the eccentricity drops enough for the collisions to become accreting. Overall, these simulations have more of the most destructive collisions (supercatastrophic, erosion) than those with lower initial eccentricities (see Figure \ref{fig:coll-freq}).

In general, in these simulations there are a wide variety of collision types. Collisions can have impact velocities of up to 20 times the escape velocity. In general, the most common collision type is hit-and-run, though in simulations with lower initial eccentricities partial accretion is the most common outcome. On average, planets are the result of 37 giant collisions during the formation process, though there is a wide spread of values, anywhere from zero up to 160 collisions per planet. More massive planets ($\geq 0.2$ M$_{\oplus}$ )  have 59 collisions on average, while small planets, unsurprisingly, form after fewer collisions, around an average of 5. We find that collisions between smaller embryos ( $ < 0.2$ M$_{\oplus}$) have different frequencies of collision types than those for larger embryos  ( $ > 0.2$ M$_{\oplus}$). Grazing impacts, like hit-and-runs, are the most frequent collisions between small embryos, and they have higher frequencies of destructive collisions, like erosion and super-catastrophic. This is good for forming Mercury analogues, as both of these collision types are thought to be capable of forming Mercury. Meanwhile, collisions between larger embryos are strongly dominated by partial accretion collisions. This is likely due to a combination of the higher escape velocity that larger embryos have, which reduces the $v_{imp}/v_{esc}$ ratio and thus results in less destructive collisions, and the fact that the larger embryos have a larger effective cross-section, which results in more head-on collisions.

\begin{figure*}
    \centering
    \includegraphics[width=0.9\textwidth]{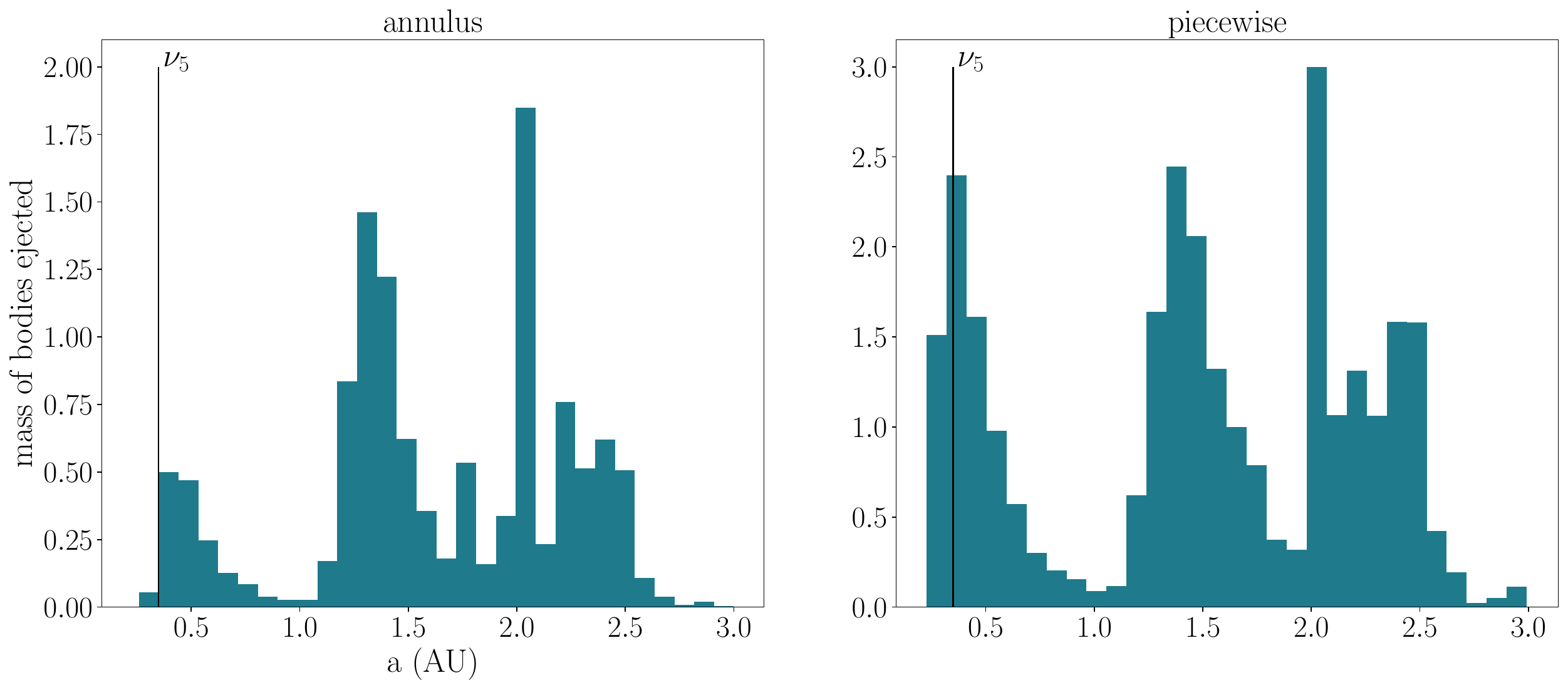}
    \caption{The total mass of bodies ejected in each semi-major axis bin for `piecewise' simulations (left) and `annulus' simulations (right). The black line shows the rough location of the $\nu_5$ resonance. There is a peak in mass loss just to the right of this line for both simulation types.}
    \label{fig:ejected}
\end{figure*}

We expected that the $\nu_5$ secular resonance with Jupiter around 0.35 AU would result in debris and other material getting absorbed by the Sun (and therefore ejected from the system), as the resonance would destabilize the bodies, thus allowing for less debris re-accretion where Mercury forms. Figure \ref{fig:ejected} shows the distribution of mass ejected from the system as a function of its semi-major axis; there is indeed a peak in mass loss near 0.35 AU,  more prominent in the `piecewise' simulations than in the `annulus' simulations due to the increased mass present at those locations. However, there is not enough mass lost near 0.35 AU to deplete the inner disks down to a few times Mercury's mass. Thus, the impact of the $\nu_5$ resonance was not as strong as expected, or there are other factors at play that replenish the mass lost at the inner edge of the disk. In both types of simulations there are also peaks in the mass ejected around 1.5 AU, and between 2 and 2.5 AU for both types of initial conditions. The outer peak is due to a number of resonances with the giant planets (i.e. $\nu_5$, $\nu_6$, and the 3:1 mean motion resonance with Jupiter \citep{1995Moons,1997Michel,2000Gladman}). The $\nu_5$ resonance is also present at higher inclinations at the peak near 1.5 AU \citep{1997Michel,2023Fenucci}. Since these disks start out with a wide range of inclinations, these resonances still have a significant effect in this area. Additionally, a number of small, eccentric embryos tend to be ejected from the main disk and end up near 1.5 AU, exciting each other and the surrounding debris. This both ejects debris and embryos on its own, and increases the number of bodies affected by the $\nu_5$ resonance at higher inclinations. Conversely, bodies excited by the $\nu_5$ resonance may then be ejected via close encounters with the embryos in the area.

\begin{figure}[h]
    \centering
    \includegraphics[width=0.5\textwidth]{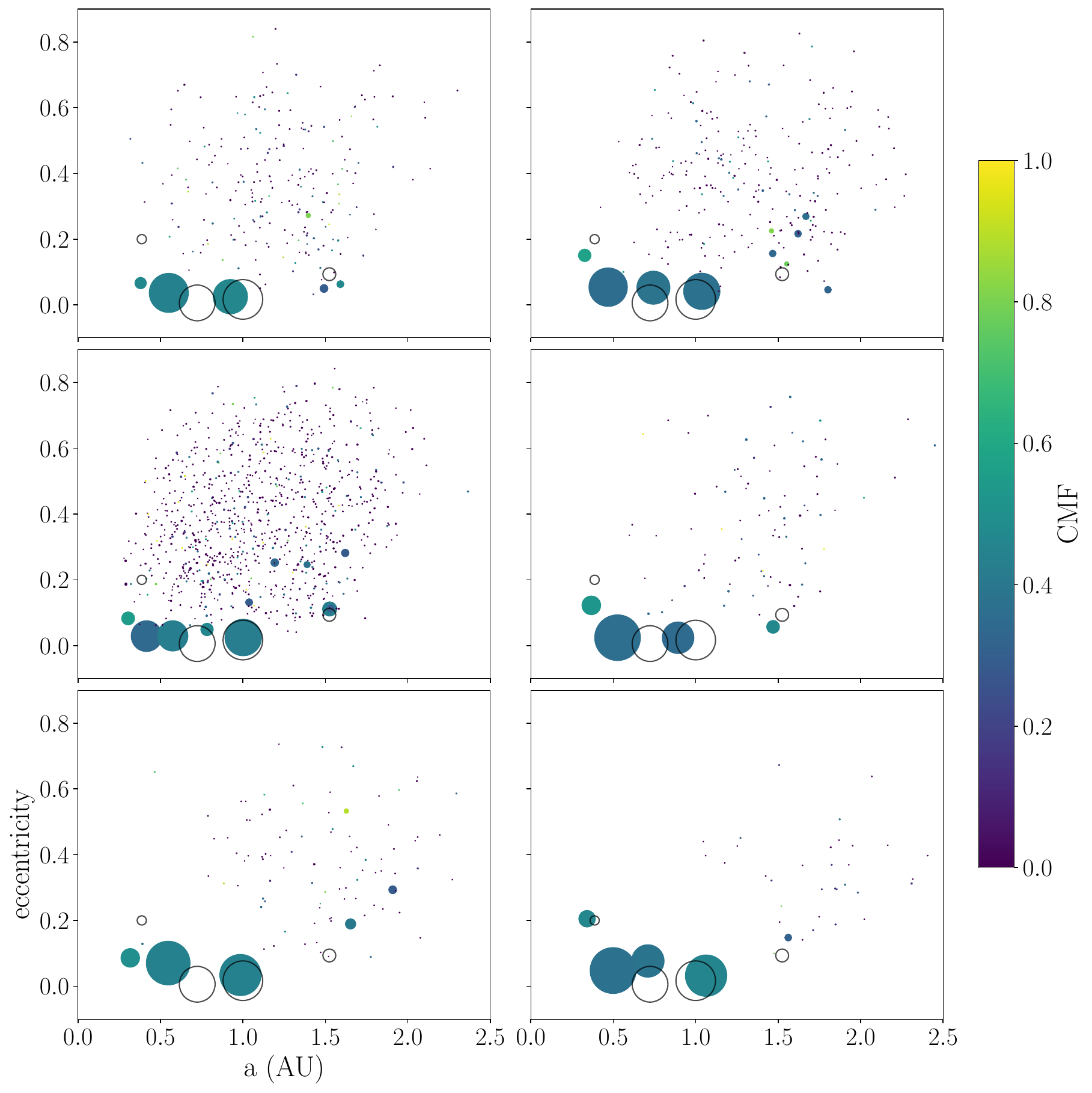}
    \caption{Systems with analogues of all four inner solar system planets and Mercury analogues with CMF $> 0.4$. The size of the bodies represents their mass, and the empty circles show the locations and masses of the actual solar system planets.}
    \label{fig:systems}
\end{figure}

\subsection{Dynamical solar system analogues}\label{subsec:sol-sys-analogues}

We use similar criteria to identify our solar system planet analogues as other studies in the field, \citep[i.e.][]{2021Clement,2019Lykawka}. The analogues in this section are identified as follows:
\begin{itemize}
    \item a Mercury analogue lies at a semi-major axis between 0.25 and 0.5 AU, and has a mass between 0.025 and 0.25 $M_{\oplus}$
    \item Earth and Venus analogues lie between 0.5 and 1.3 AU, and are above 0.6 $M_{\oplus}$. If there is one analogue in this space it is an Earth analogue, and a second analogue is the Venus analogue
    \item a Mars analogue lies between 1.3 and 2 AU, and is less than 0.3 $M_{\oplus}$
\end{itemize}

Out of our 117 simulations, $54\%$ have at least three solar system analogues, and only $11\%$ have analogues of all four solar system planets. Figure \ref{fig:systems} shows a few of the best solar system analogues out of our simulated systems. Most synthetic solar systems have additional mass outside of the solar system planet analogues, both in additional planets and also in debris. The average number of planets per simulation is 6.5, higher than the four we aim for. Typically, these extra planets are Mars analogues. In these simulations, we consider a planet to be any body that is above the minimum embryo mass in the simulation, which is $2.5 \times 10^{-3} M_{\oplus}$. The excess total mass in our results compared to that of the terrestrial planets is due to the large initial mass we assumed, but this can presumably be adjusted by starting with a less massive disk.

\begin{figure*}
    \centering
    \includegraphics[width=0.9\textwidth]{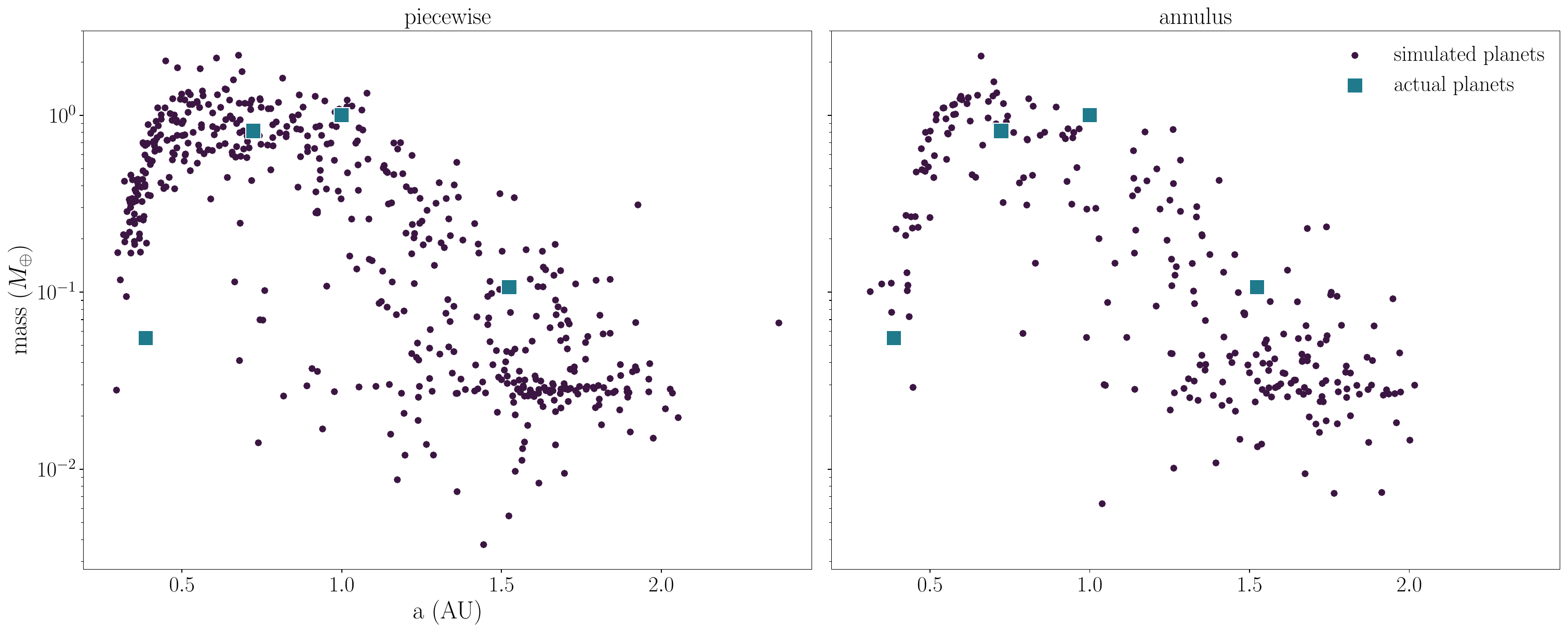}
    \caption{The masses and semi-major axes of all planets formed in our simulations, compared to the mass and semi-major axis of the four inner terrestrial planets. The distribution mostly matches the four planets for both disk types, though the inner planets are slightly too massive to include Mercury for the `piecewise' disks.} 
    \label{fig:all-pls}
\end{figure*}

Figure \ref{fig:all-pls} shows the mass and semi-major axis distribution of planets across all of our simulations, compared to the solar system planets. Overall, we approximately reproduce the distribution, similar to previous works, and more importantly we achieve the appropriate masses and semi-major axes for Mercury analogs. Just over $26\%$ of systems have a Mercury analog in mass and semi-major axis. There is a significant tail of planets near Mars, which is due to the fact that simulations tend to have multiple Mars-analogues in a system instead of just one. However, these often also have high eccentricities, and so may be ejected given long enough simulation time. We ran a few simulations for longer in order to test their long-term stability. In these runs, Mars analogues tended to decrease in number, particularly in runs with many. In addition, $\sim 15\%$ of the dynamical Mercury analogues in those simulations were lost. This adds further weight to the idea of unstable Mercury analogues as discussed in Section \ref{subsec:proto-mercuries}. However, due to the length of time required to continue these simulations beyond 100 Myr, a full analysis of the long-term stability of these planetary systems is outside the scope of this work.  

The distribution in Figure \ref{fig:all-pls} also skews towards higher masses than expected for the solar system, and at smaller semi-major axes. In particular, Venus analogues tend to be more massive and closer to the Sun than expected. This is in large part due to an excess of initial mass in the inner disk portion, and is therefore more pronounced in runs with the initial `piecewise' disk distribution. However, those with the typical `annulus' also have Venus and Earth analogues that are too close to the Sun, as seen in Figure \ref{fig:all-pls}, indicating that the initial distribution does not completely account for the mismatch. We discuss this further in Section \ref{subsec:giantpls}. 

{An additional result of the Venus analogues forming too close to the Sun is that Venus and Mercury tend to form too close together.  This is a consistent issue in formation simulations of the solar system, as mentioned in Section \ref{sec:intro}, and our systems are no exception. The majority of systems have  period ratios of Venus analogues to Mercury analogues below the solar system value of 2.55. A handful of systems have a similar value to the solar system, and some have an even higher value. However, those with a higher period ratio also mostly have an additional body that has formed between Venus and Mercury.

\subsection{Mercury analogues}\label{subsec:mercury-analogues}

\begin{figure*}
    \centering
    \includegraphics[width=0.9\textwidth]{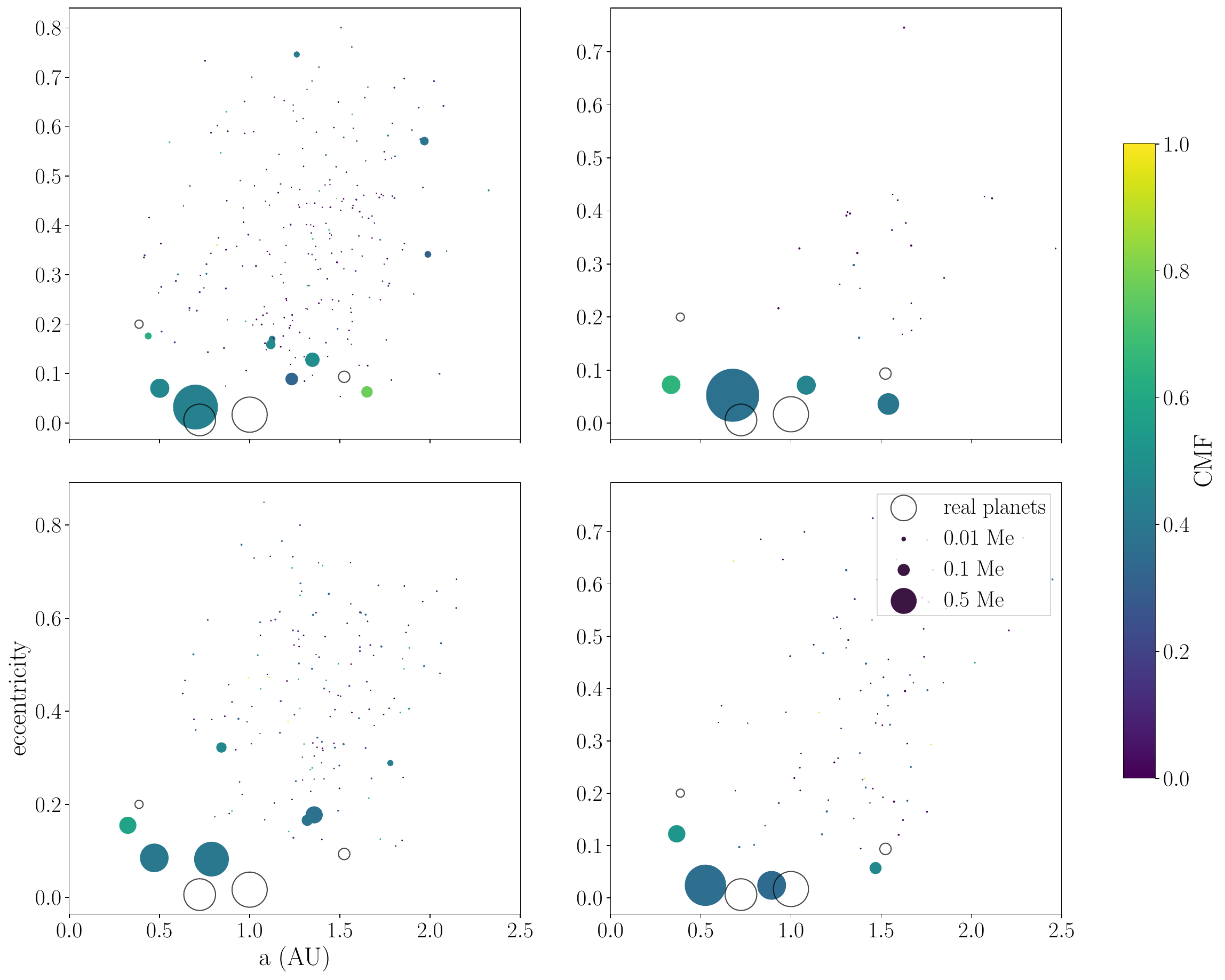}
    \caption{The systems with Mercury analogues with CMFs $> 0.5$. Top left is an `annulus' simulation, and the rest are `piecewise' simulations. Two are from simulations with initial average eccentricity of 0.3. The solar system planets are represented by empty circles for comparison.}
    \label{fig:mercuries}
\end{figure*}

Our dynamical criteria for a Mercury analogue is described above, in Section \ref{subsec:sol-sys-analogues}. We also consider various minimum CMFs for a Mercury analogue, since Mercury's CMF is thought to be $\sim 0.7$ \citep{2013Hauck}. Figure \ref{fig:mercuries} shows all surviving Mercury analogues with a CMF above 0.4. This reduces the Mercury occurrence rate from $26\%$ to $10\%$. These analogues tend to be more massive than Mercury, and are mostly found between 0.3 AU and Mercury's orbit (0.38 AU). Only four of these Mercuries have a CMF above 0.5 (plotted above in Figure \ref{fig:mercuries}), so increasing the minimum CMF to 0.5 drops the Mercury occurrence rate to $\sim 3\%$. We do not get any Mercury analogues with a CMF of $\sim 0.7$.  One additional constraint we consider is the difference between the CMF of Earth and Mercury. In the solar system, Earth's CMF is $\sim 0.326$ \citep{2005Stacey}, less than half of Mercury's CMF.  Only two of our simulations have both a Mercury and an Earth analogue where the Mercury's CMF is 1.5 times that of the Earth's, and both of them are simulations where the Mercury analogue's CMF is $> 0.5$. 

Mercuries (with a CMF $> 0.4$) form with a slightly higher occurrence rate in the `piecewise' simulations ($11\%$) than in the `annulus' simulations ($8\%$), though this difference is so small that it is statistically insignificant. However, three of the four Mercuries with CMF $> 0.5$ are from `piecewise' simulations, which suggests that these simulations are better at forming Mercuries with higher CMFs. Mercury analogues also seem evenly spread across simulations with different initial eccentricities and inclinations. They form more frequently in simulations with $50\%$ debris loss and therefore higher disk masses. On average, these Mercury analogues have eccentricities of 0.15. 

Mercury analogues have a variety of histories. The two smallest formed after having been the second largest remnant in two or more hit-and-run collisions. This is one of the most prevalent theories for Mercury's collisional history \citep{2014Asphaug,2018Chau,2018Jackson}. Thus, this does seem to be the best way to form a high CMF, low-mass Mercury, though it does not appear to be very common in our simulations. The other Mercuries form from a variety of collisions, but typically at least one erosion collision, and often a few hit-and-runs, although they are often the largest remnant in these collisions.

\subsection{Proto-Mercury analogues}\label{subsec:proto-mercuries}

In order to get a better sense of how Mercuries form, and their survival rate, we track the formation and fates of proto-Mercuries across our simulations. The criteria we set for a body to be a proto-Mercury follows the dynamical criteria for a Mercury analogue as described in Section \ref{subsec:sol-sys-analogues}. We require a minimum CMF of 0.4, as in Section \ref{subsec:mercury-analogues}. Finally, we place some limits on the duration of the proto-Mercury's existence (i.e. the length of time that it meets the above criteria), and the time by which the proto-Mercury has formed. For the former, we require it to exist for a minimum of 5 Myr. For the latter, we require that the proto-Mercury exists a minimum of 10 Myr after the start of the simulation, or once the largest embryo in the simulation reaches: $m_{max} = 0.5 M_{\oplus} + m_{init}$, where  $m_{init}$ is the initial mass of the embryos in that simulation. This is a rough approximation that the simulation has reached or is nearing the oligarchic phase, where more massive embryos dominate the evolution of the disk. This also typically corresponds to a time of $10 - 20$ Myrs, which is roughly the time by which most Mercury analogues have reached their final semi-major axis.

Including surviving Mercury analogues, 25\% of our simulations have at least one proto-Mercury, and $\sim 4\%$ of the simulations have two or more proto-Mercuries that form. We also consider `dynamical' proto-Mercuries with no minimum CMF (criteria as described in Section \ref{subsec:sol-sys-analogues}), and find that these form in 40\% of simulations, where there are 2 or more proto-Mercuries in 10\% of simulations. Both of these are roughly twice the occurrence rates of surviving Mercuries in our simulations. 
Thus, forming Mercury-like bodies is easier than keeping them in the system. 

The majority of proto-Mercuries that do not survive are removed from the system after being engulfed by another body. This other body is usually either another proto-Mercury or a Venus analogue orbiting further out, which in our simulations is at $\sim 0.5$ AU. After engulfment, the other ways of `losing' a proto-Mercury include it exceeding the mass threshold, falling below the CMF threshold, and moving out of the semi-major axis limits ($a > 0.5$ AU). The first two are the most common. In some cases, the proto-Mercury will stop being a Mercury shortly before being absorbed by another body. The problem does not seem to be forming a Mercury-like body, but to keep it in a stable orbit and at a high CMF. 

There are some differences between Mercuries in `annulus' disks versus those in `piecewise' disks. First of all, dynamical proto-Mercuries (with no minimum CMF) are significantly more frequent in `annulus' simulations compared to `piecewise' simulations. As stated above in Section \ref{subsec:mercury-analogues}, the opposite is true for surviving Mercury analogues (though by a small amount). Thus, it may be easier to form Mercuries with the `annulus' initial conditions, but those Mercuries are also much more unstable. It makes sense that these proto-Mercuries would be unstable, because the initial disk does not extend into the range of semi-major axes allowed for Mercury, so these embryos \textit{have} to be knocked inwards by close encounters or a collision. However, this does not mean that all Mercury analogues in the `piecewise' disks start between 0.25 and 0.5 AU. Often, `piecewise' Mercury analogues move into this range from further out in the disk, although the process is less abrupt than in the `annulus' disks as there is a continuous distribution of material across much of these semi-major axes. An additional feature of the `piecewise' proto-Mercuries is that they have a higher probability of becoming too massive, since in these simulations the disk starts with more material in the Mercury semi-major axes, allowing for more accretion of small embryos and debris.

\section{Discussion}\label{sec:disc}

\subsection{Adjusting debris mass loss}\label{subsec:adj-debmass}

As discussed above in Section \ref{subsec:mercury-analogues}, our highest CMF Mercury analogues have CMFs $< 0.7$, Mercury's inferred CMF \citep{2013Hauck}. The planets reach CMFs of 0.7 or higher during their formation, but other collisions and debris accretion in particular result in a decrease of the final planet's CMF, even considering that we remove 25-50\% of the debris mass formed at every collision due to collisional grinding. However, recent work suggests that the fraction of debris lost to collisional grinding in the solar system may be significantly higher, as high as 98\%, in order to avoid over-populating the asteroid belt with debris from collisions (Admane and Jacobson, in prep). A higher debris mass loss fraction may also have a significant impact on the final CMFs of our Mercury analogues, so here we test what fraction of the debris mass has to be removed in order to elevate our Mercury analogues' CMFs to 0.7. 

\begin{figure*}
    \centering
    \plottwo{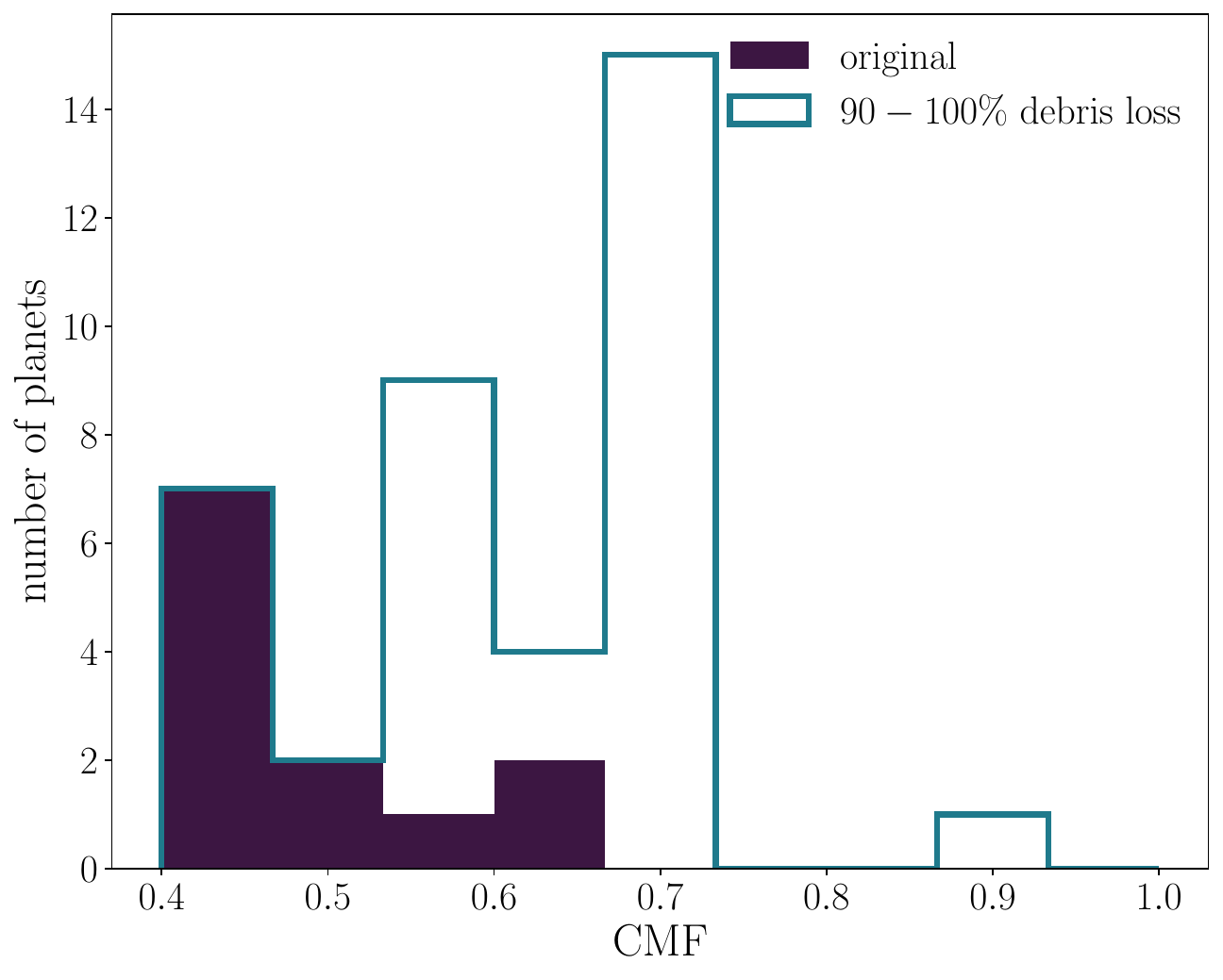}{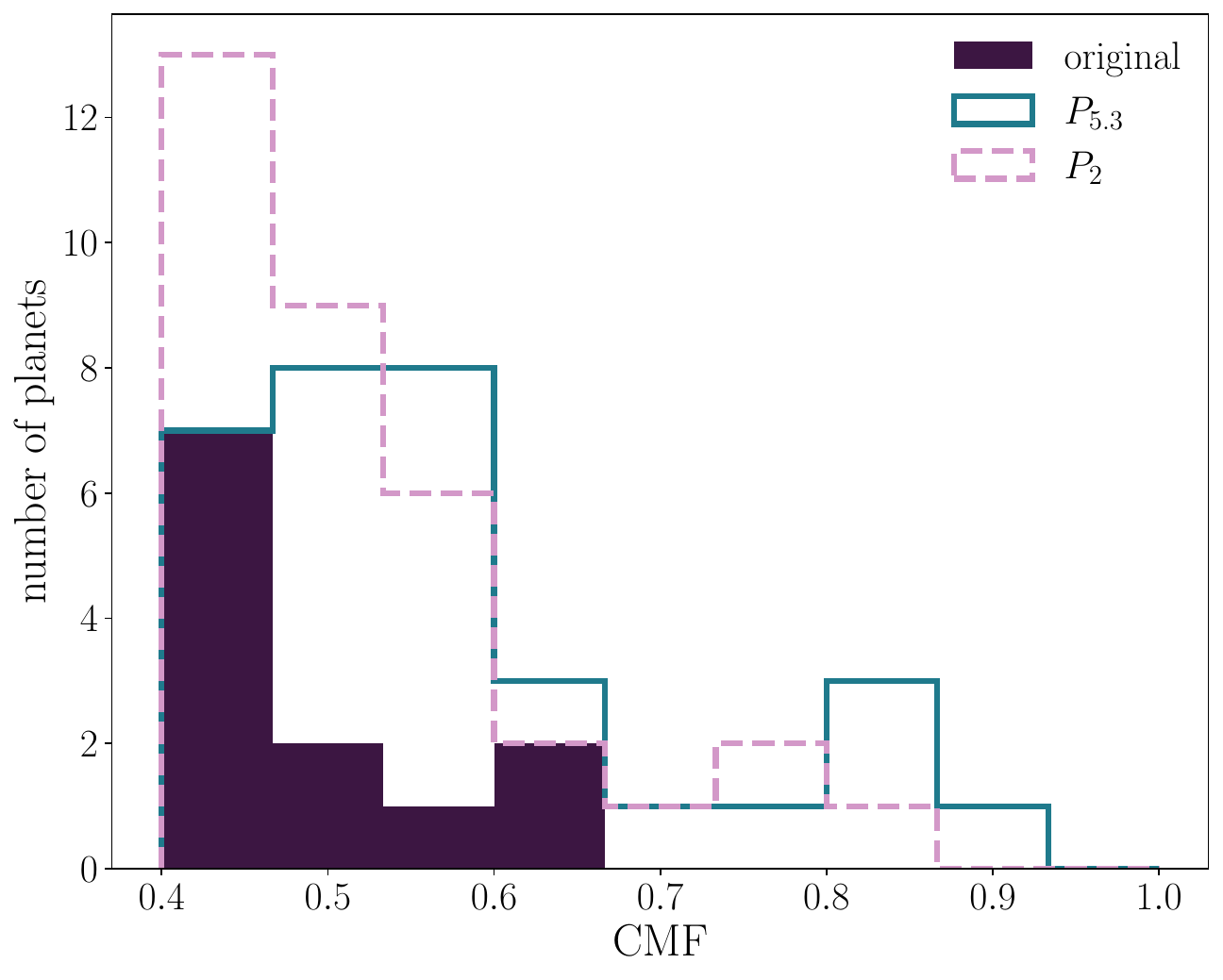}
    \caption{The distribution of Mercury analogues in the original simulations (solid purple), compared to Mercury analogues found when debris mass loss rates are 90-100\% (left) and when debris mass loss follows two different power laws (right).}
    \label{fig:merc-comparison-hist}
\end{figure*}

We calculated the fraction of the final planet's mass that came from debris, and the CMF  of the debris that was accreted. Assuming that the CMF of the accreted debris remains the same, we then calculate how much debris mass has to be removed in order to increase the CMF of the final planet as close to 0.7 as possible. Roughly half of the Mercuries cannot reach a CMF of 0.7 even with removal of 100\% of the debris mass they accreted. The other half can reach 0.7 CMF with an adjusted total debris mass loss spread from 60 - 98\%. We also perform this calculation for those proto-Mercuries that became too massive, or their CMF dropped too low, to see if this calculation would recover any of these to the point that they could be considered Mercury analogues again. We in fact find that the majority of the proto-Mercuries in these categories become Mercury analogues by our criteria set out in Sections \ref{subsec:sol-sys-analogues} and \ref{subsec:mercury-analogues} with 90-100\% total debris mass loss. Many of them do not reach a CMF of 0.7, but most of them have CMF $\geq 0.4$. Additionally, those proto-Mercuries that became too massive, when they lose most of the debris that was accreted, return to masses $< 0.25 M_{\oplus}$. With this higher rate of proto-Mercury survival, we get more than double the number of remaining Mercury analogues,  an occurrence rate of $34\%$ (see Figure \ref{fig:merc-comparison-hist}). 

This is not a perfect solution, unfortunately. We also calculate the new CMFs of the other planets in the system with this new debris mass loss fraction that is required for the Mercuries to have a CMF of 0.7. While some planets retain their initial CMF, many of the planets in the system have accreted a large fraction of their mass from debris, and so the removal of this debris significantly increases their CMFs as well. More massive planets, such as the Venus and Earth analogues, can have CMFs as high as 0.8 or 0.9. Since the Earth's CMF is known to be $0.326$, and Venus' is thought to be similar \citep{2005Stacey}, these high debris mass loss scenarios do not match the constraints of the whole solar system. Thus, with this simple scaling, it seems unlikely that a higher overall fraction of debris removal in our simulations would have provided accurate Earth and Venus analogues, even if it did improve the Mercury analogues and the rate of Mercury analogues. 

{To that end, we considered the possibility that the debris mass loss fraction varies with respect to the planet's semi-major axis. \citet{Watt2024} simulated the evolution of debris from giant impacts at different semi-major axes, showing that giant impacts that occur at smaller semi-major axes produce more massive debris disks via collisional grinding. Additionally, \citet{2020Spalding} showed that the stronger solar winds likely present at the time of Mercury's formation should result in increased outward migration of debris particles, faster closer in to the Sun and slower further out. It follows that the debris mass loss fraction would be higher for closer-in planets like Mercury and lower for planets further from the Sun due to the combination of these two effects. To test how this would influence the creation of Mercury analogues, we constructed two power laws to describe debris mass loss fraction over semi major axis. One was loosely fitted to the mass of the debris disks created at different semi-major axes in \citet{Watt2024} (which we will call $P_{5.3}$ \footnote{The equation used for this was: $y = 1.99\times10^{-3}x^{-5.3}$}), and for the other we chose a simple power law of $x^{-2}$ (which we will call $P_{2}$) \footnote{The equation used for this was $y=6\times10^{-2}x^{-2}$}. Both power laws were normalized such that they reach a maximum debris mass loss fraction of 1 around 0.3 AU, but drop off after 0.5 AU.  We then used both power laws to recalculate any increased debris mass loss for each planet based on its final semi-major axis.

With the newly recalculated debris mass loss fraction for each planet, we calculate a corresponding new mass and CMF. As might be expected, the change in planet CMF follows a similar slope to that of the debris mass loss power laws we used, increasing exponentially for planets interior to 0.4 or 0.5 AU. Unlike the previous method, this means that the analogues of the other planets remain relatively unchanged. Similar to above, we take this updated set of planet masses and CMFs and search for Mercury analogues. The comparison between the original distribution of Mercury analogues and those existing in the new sample is shown in Figure \ref{fig:merc-comparison-hist}. It is clear from this figure that there is a significant increase in Mercury analogues for both power laws. For $P_{5.3}$ the rate of Mercury analogues is $\sim 27\%$, similar to when an overall debris mass loss rate of $90-100\%$ is considered. For $P_2$, the rate of Mercuries similar at $\sim 29\%$. Both are a significant increase from the $10\%$ occurrence rate we find in the original simulations. There remains the caveat that this is just an approximation, as we assume a debris mass loss rate based on a planet's final location, instead of its location over time. Additionally, removing debris can change the dynamical evolution in unanticipated ways. Thus, we suggest that incorporating a debris mass loss fraction that changes with semi-major axis for giant collisions could have a significant impact on the creation of Mercury analogues in future solar system simulations, and is an important avenue for future work.

\subsection{Debris accretion by inner planets}\label{subsec:giantpls}

In this study, we assume an early instability, so the giant planets start in their current positions and have their current orbital parameters. We chose this configuration so that the initial disk of embryos would start excited by this instability, and so the giant planets would be in their current configurations, placing the $\nu_5$ secular resonance with Jupiter near Mercury's orbit. \citet{2021Clement} found that this resonance, in combination with others, destabilized proto-Mercuries just inside of Mercury's current orbit, around 0.35 AU. Thus, we hypothesized that the $\nu_5$ resonance would similarly destabilize and eject into the Sun much of the debris created in collisions around Mercury. This would then result in reduced debris accretion near Mercury, and possibly a higher CMF for Mercury analogues. 

As mentioned in Section \ref{subsec:evolution}, Figure \ref{fig:ejected} shows that there is in fact mass loss around the inner $\nu_5$ resonance, more prominent for simulations with a `piecewise' initial condition because there is more mass near the resonance to begin with. Therefore, the $\nu_5$ resonance seems to be actively destabilizing debris mass that then leads to its ejection. However, we do not find that the interior planets have less debris re-accretion than other planets. In fact, we find that the average fraction of a planet's mass that was accreted in debris throughout its formation period increases as its semi-major axis decreases, with the planets forming around 0.35 AU (also the closest planets to the Sun that form) having the highest mass fraction of debris. This is true for planets with all initial conditions, including the `annulus' initial condition that starts without any mass interior to 0.7 AU. 

This can be attributed to a combination of factors. The first is that the Safronov number of the planets decreases as their semi-major axes decrease. The Safronov number is given by: $\Theta = \frac{v^2_{esc}}{2 v^2_{orb}}$, where $v_{orb}$ is the orbital velocity of the object \citep{1972Safronov}. It determines the fate of a planetary system where there is no significant source of damping; where this number is greater than one, close encounters tend to lead to scattering, and where it is less than one, they tend to lead to collisions \citep{2018Exoplanets}. This therefore means that close encounters between planets and debris are more likely to be accretionary the closer they are to the Sun. Secondly, almost half of all debris is created inside of 0.7 AU, despite the fact that all simulations begin with much less than half of their mass within 0.7 AU. There is therefore a disproportionate amount of debris in the inner disk that is then available to be accreted by the interior planets. This excess of debris can be explained by the fact that in disks of embryos with high eccentricities, the impact probability increases as the distance from the Sun decreases \citep{2003Levison}, and thus collisions and debris creation happen more frequently closer to the Sun. This also explains why our simulations end up with too much mass close to the Sun, where the Mercury analogues form, despite our initial distributions placing most of the mass between 0.7 and 1 AU. It is also possible that some of the debris destabilized by the $\nu_5$ resonance collides with inner planets instead of the Sun, though it seems like this would be a minor effect. 

Thus, in order to better form a high-CMF Mercury, some other mechanism may be needed to more efficiently remove debris. As discussed in the Section \ref{subsec:otherwork} below, an early instability could sweep the $\nu_5$ resonance inwards through the disk, which should force more debris to be thrown towards the Sun and thus leave less to be accreted by Mercury.  It is also possible that the $\nu_5$ resonance could increase the fraction of mass loss due to collisional grinding for debris formed near this resonance, as it would bump up the eccentricities of the debris. As well, collisional grinding may be more efficient closer to the Sun, as \citet{2010Kobayashi} found that the mass depletion time for collisional cascade increases with decreasing semi-major axis. This would help to remove some of the excess debris, allowing Mercury analogues to reach higher CMFs and smaller masses. Finally, a scenario where Jupiter and Saturn start out with more eccentric orbits than their current ones would increase the strength of the secular resonances, possibly resulting in more material being removed by the inner $\nu_5$ resonance \citep{2009Raymond}.

\subsection{Comparison to other works}\label{subsec:otherwork}

As mentioned in Section \ref{sec:intro}, some recent works have also had some success with forming Mercury analogues.  \citet{2021bClement} formed Mercuries at a $10\%$ occurrence rate with an inner disk of material, similar to the `piecewise' initial conditions used in this study. However, they placed Jupiter and Saturn in mean-motion resonance, on their pre-instability orbits, and tested different slopes and distributions. They also found that they formed high-CMF ( $> 0.5$ CMF) Mercury analogues around 20\% of the time, which is a significantly higher rate than in our results. It seems that the treatment of collisions in the \texttt{MERCURY} code may be the cause. Unlike in \texttt{SyMBA}, \texttt{MERCURY} creates debris particles that are treated as embryos after the collision, and thus the `minimum fragment mass' is orders of magnitude larger for \texttt{MERCURY} (0.0055 $M_{\oplus}$) than it is in this study ($1.5\times 10^{-5} M_{\oplus}$). With such large fragments, each one can make a significant difference in the CMF of the planet it accretes on. More importantly, if only one fragment is not re-accreted, this keeps the CMF of the planet much higher than it would be if only one debris particle was not re-accreted in the simulations in this work. Thus, we suggest that it is possible that the high Mercury CMFs in \citet{2021bClement}, for example, may result from the way that debris is treated. It is not the only possible explanation, due to the other differences in the simulations of \citet{2021bClement} (i.e. giant planet positions). Even so, this highlights the importance of ensuring that the debris from collisions is treated as realistically as possible in planet formation simulations. 

Additionally, \citet{Clement2023} found that they form Mercury analogues frequently in their simulations, using a similar initial disk as above, where they also simulate the early giant planet instability and its effects on the inner terrestrial disk. In these simulations, the sweeping of the $\nu_5$ resonance inwards over the course of the formation, caused by the instability, seemed to remove enough material in the Mercury region to form better Mercury analogues. Thus, it is possible that including such a process in our simulations may have had a similar effect, and may have reduced the debris accretion onto Mercury analogues.

\section{Summary and Conclusions}\label{sec:conc}

In this study, we simulate the formation of the solar system with a focus on the formation of a high-CMF Mercury. In order to facilitate Mercury's formation, we start with an inner disk of material from 0.3 to 0.7 AU, attached to the typical annulus of embryos from 0.7 to 1 AU, and we start with the giant planets in their current configurations. We find that $11\%$ of simulations have an analogue for each of the four inner terrestrial planets. Overall, we form too many planets per system, and specifically too many planets with high mass near 0.5 AU. This is due to initial disk masses that are too large, and particularly overly massive inner disks. These disk masses were adjusted for mass loss from the collisional grinding of debris, but less mass loss occurred than anticipated. 

Mercury analogues with CMF $\ge$ 0.4 form in $\sim 10\%$ of simulations, at a similar rate to recent studies with similar initial conditions \citep{2021bClement}. However, we fail to form Mercuries with CMFs $\ge$ 0.69, which is the minimum estimate of Mercury's CMF \citep{2013Hauck}. It seems that this is due to an excess of mass and debris accretion at the inner edge of the disk, and thus one solution could be that the giant planet instability happens during the formation period, as in \citet{Clement2023}, causing the $\nu_5$ resonance to sweep through the inner disk and push debris towards the Sun. However, as our Mercury analogues seem to have lower CMF than those of \citet{2021bClement}, we also find that the implementation of debris from collisions in \texttt{MERCURY} may not be high-resolution enough to emulate the re-accretion of debris. 

More generally, we draw the following conclusions from the simulations we performed:
\begin{enumerate}
    \item More proto-Mercuries form than remain in the final systems. Typically, they are removed via accretion onto other bodies in the system. This suggests that one reason for Mercury's lower occurrence rates in simulations may simply be that its formation pathway is inherently more unstable than the other planets. 
    \item Debris accretion accounts for a significant mass fraction of most planets formed, including Mercury analogues. In our work, this means that high-CMF Mercury analogues form most commonly in simulations with higher (50\% instead of 25\%) debris mass loss. This highlights the importance of improving our understanding of what happens to debris post-collision, and subsequently applying those improvements to our implementations of debris creation in planet formation simulations. 
\end{enumerate}

Our conclusions demonstrate that the most important areas of improvement for Mercury's formation still lie in the details of the collisions that form it. Additionally, they suggest that parameters that may increase the formation rate of high-CMF Mercuries may also result in lower-quality Earth analogues. Thus, a cohesive picture of the formation of all the inner planets and their compositions requires more thorough implementations of collisions and the debris they produce.

\begin{acknowledgments}
JS and DV are supported by the Natural Sciences and Engineering Research Council of Canada (grant RGPIN-2021-02706). AM is grateful for support from the ERC advanced grant HolyEarth N. 101019380. We thank the anonymous reviewer for their comments which helped improve the manuscript. JS would like to thank Norman Murray and Yanqin Wu for helpful discussions about the initial conditions and analysis of the simulations that improved the quality of this work. The simulations were performed on the Sunnyvale computer at the Canadian Institute for Theoretical Astrophysics (CITA) and the facilities of the Shared Hierarchical 
Academic Research Computing Network (SHARCNET:www.sharcnet.ca) and Compute/Calcul Canada.
This research was enabled in part by support provided by Compute Ontario (www.computeontario.ca) and the Digital Research Alliance of Canada (alliancecan.ca).
We would like to acknowledge that our work was performed on land traditionally inhabited by the Wendat, the Anishnaabeg, Haudenosaunee, Metis and the Mississaugas of the New Credit First Nation.  
\end{acknowledgments}

\vspace{5mm}


\software{\texttt{astropy} \citep{astropy:2013,astropy:2018}, \texttt{SyMBA} \citep{Duncan1998}
          }







\bibliography{forming-mercury}{}
\bibliographystyle{aasjournal}



\end{document}